\documentclass[reprint,superscriptaddress,amsmath,amssymb,aps,prc,twocolumn]{revtex4-2}

\usepackage{graphicx}
\usepackage{hyperref}
\usepackage{color}
\usepackage[english]{babel}
\usepackage{bm}
\usepackage[T1]{fontenc}
\usepackage[output-decimal-marker={.},exponent-product=\cdot]{siunitx}

\begin{document}

\title{Isomeric states of fission fragments explored via Penning trap mass spectrometry at IGISOL}

\author{A.~Jaries}
\email{arthur.a.jaries@jyu.fi}
\affiliation{University of Jyvaskyla, Department of Physics, Accelerator Laboratory, P.O. Box 35(YFL) FI-40014 University of Jyvaskyla, Finland}
\author{M. Stryjczyk}
\email{marek.m.stryjczyk@jyu.fi}
\affiliation{University of Jyvaskyla, Department of Physics, Accelerator Laboratory, P.O. Box 35(YFL) FI-40014 University of Jyvaskyla, Finland} 
\author{A.~Kankainen}
\email{anu.kankainen@jyu.fi}
\affiliation{University of Jyvaskyla, Department of Physics, Accelerator Laboratory, P.O. Box 35(YFL) FI-40014 University of Jyvaskyla, Finland}
\author{L.~Al~Ayoubi}
\affiliation{University of Jyvaskyla, Department of Physics, Accelerator Laboratory, P.O. Box 35(YFL) FI-40014 University of Jyvaskyla, Finland}
\affiliation{Universit\'e Paris Saclay, CNRS/IN2P3, IJCLab, 91405 Orsay, France}
\author{O.~Beliuskina}
\affiliation{University of Jyvaskyla, Department of Physics, Accelerator Laboratory, P.O. Box 35(YFL) FI-40014 University of Jyvaskyla, Finland}
\author{L.~Canete}
\altaffiliation[Present address: ]{Northeastern University London, Devon House, 58 St Katharine's Way, E1W 1LP, London, United Kingdom}
\affiliation{University of Jyvaskyla, Department of Physics, Accelerator Laboratory, P.O. Box 35(YFL) FI-40014 University of Jyvaskyla, Finland}
\author{R.P.~de Groote}
\altaffiliation[Present address: ]{KU Leuven, Instituut voor Kern- en Stralingsfysica, B-3001 Leuven, Belgium}
\affiliation{University of Jyvaskyla, Department of Physics, Accelerator Laboratory, P.O. Box 35(YFL) FI-40014 University of Jyvaskyla, Finland}
\author{C.~Delafosse}
\altaffiliation[Present address: ]{Universit\'e Paris Saclay, CNRS/IN2P3, IJCLab, 91405 Orsay, France}
\affiliation{University of Jyvaskyla, Department of Physics, Accelerator Laboratory, P.O. Box 35(YFL) FI-40014 University of Jyvaskyla, Finland}
\author{P.~Delahaye}
\affiliation{GANIL, CEA/DSM-CNRS/IN2P3, Boulevard Henri Becquerel, 14000 Caen, France}
\author{T.~Eronen}
\affiliation{University of Jyvaskyla, Department of Physics, Accelerator Laboratory, P.O. Box 35(YFL) FI-40014 University of Jyvaskyla, Finland}
\author{M.~Flayol}
\affiliation{Universit\'e de Bordeaux, CNRS/IN2P3, LP2I Bordeaux, UMR 5797, F-33170 Gradignan, France}
\author{Z.~Ge}
\affiliation{GSI Helmholtzzentrum für Schwerionenforschung, 64291 Darmstadt, Germany}
\affiliation{University of Jyvaskyla, Department of Physics, Accelerator Laboratory, P.O. Box 35(YFL) FI-40014 University of Jyvaskyla, Finland}
\author{S.~Geldhof}
\altaffiliation[Present address: ]{GANIL, CEA/DRF-CNRS/IN2P3, B.P. 55027, 14076 Caen, France}
\affiliation{University of Jyvaskyla, Department of Physics, Accelerator Laboratory, P.O. Box 35(YFL) FI-40014 University of Jyvaskyla, Finland}
\author{W.~Gins}
\affiliation{University of Jyvaskyla, Department of Physics, Accelerator Laboratory, P.O. Box 35(YFL) FI-40014 University of Jyvaskyla, Finland}
\author{M.~Hukkanen}
\affiliation{University of Jyvaskyla, Department of Physics, Accelerator Laboratory, P.O. Box 35(YFL) FI-40014 University of Jyvaskyla, Finland}
\affiliation{Universit\'e de Bordeaux, CNRS/IN2P3, LP2I Bordeaux, UMR 5797, F-33170 Gradignan, France}
\author{P.~Imgram}
\altaffiliation[Present address: ]{KU Leuven, Instituut voor Kern- en Stralingsfysica, B-3001 Leuven, Belgium}
\affiliation{Institut f\"ur Kernphysik, Technische Universit\"at Darmstadt, 64289 Darmstadt, Germany}
\author{D.~Kahl}
\altaffiliation[Present address: ]{Facility for Rare Isotope Beams, Michigan State University, East Lansing, MI, USA}
\affiliation{Extreme Light Infrastructure – Nuclear Physics, Horia Hulubei National Institute for R\&D in Physics and Nuclear Engineering (IFIN-HH), 077125 Bucharest-M\u{a}gurele, Romania}
\author{J.~Kostensalo}
\affiliation{Natural Resources Institute Finland, FI-80100 Joensuu, Finland}
\author{S.~Kujanp\"a\"a}
\affiliation{University of Jyvaskyla, Department of Physics, Accelerator Laboratory, P.O. Box 35(YFL) FI-40014 University of Jyvaskyla, Finland} 
\author{D.~Kumar}
\affiliation{GSI Helmholtzzentrum für Schwerionenforschung, 64291 Darmstadt, Germany}
\author{I.D.~Moore}
\affiliation{University of Jyvaskyla, Department of Physics, Accelerator Laboratory, P.O. Box 35(YFL) FI-40014 University of Jyvaskyla, Finland}
\author{M.~Mougeot}
\affiliation{University of Jyvaskyla, Department of Physics, Accelerator Laboratory, P.O. Box 35(YFL) FI-40014 University of Jyvaskyla, Finland} 
\author{D.A.~Nesterenko}
\affiliation{University of Jyvaskyla, Department of Physics, Accelerator Laboratory, P.O. Box 35(YFL) FI-40014 University of Jyvaskyla, Finland} 
\author{S.~Nikas}
\affiliation{University of Jyvaskyla, Department of Physics, Accelerator Laboratory, P.O. Box 35(YFL) FI-40014 University of Jyvaskyla, Finland} 
\author{D. Patel}
\affiliation{Department of Physics, Indian Institute of Technology Roorkee, Roorkee 247667, India}
\author{H.~Penttil\"a}
\affiliation{University of Jyvaskyla, Department of Physics, Accelerator Laboratory, P.O. Box 35(YFL) FI-40014 University of Jyvaskyla, Finland}
\author{D.~Pitman-Weymouth}
\affiliation{Department of Physics and Astronomy, University of Manchester, Manchester M13 9PL, United Kingdom}
\author{I.~Pohjalainen}
\affiliation{University of Jyvaskyla, Department of Physics, Accelerator Laboratory, P.O. Box 35(YFL) FI-40014 University of Jyvaskyla, Finland}
\author{A.~Raggio}
\affiliation{University of Jyvaskyla, Department of Physics, Accelerator Laboratory, P.O. Box 35(YFL) FI-40014 University of Jyvaskyla, Finland}
\author{M.~Ramalho}
\affiliation{University of Jyvaskyla, Department of Physics, Accelerator Laboratory, P.O. Box 35(YFL) FI-40014 University of Jyvaskyla, Finland}
\author{M.~Reponen}
\affiliation{University of Jyvaskyla, Department of Physics, Accelerator Laboratory, P.O. Box 35(YFL) FI-40014 University of Jyvaskyla, Finland}
\author{S.~Rinta-Antila}
\affiliation{University of Jyvaskyla, Department of Physics, Accelerator Laboratory, P.O. Box 35(YFL) FI-40014 University of Jyvaskyla, Finland}
\author{A.~de~Roubin}
\affiliation{University of Jyvaskyla, Department of Physics, Accelerator Laboratory, P.O. Box 35(YFL) FI-40014 University of Jyvaskyla, Finland}
\affiliation{Universit\'e de Bordeaux, CNRS/IN2P3, LP2I Bordeaux, UMR 5797, F-33170 Gradignan, France}
\author{J.~Ruotsalainen}
\affiliation{University of Jyvaskyla, Department of Physics, Accelerator Laboratory, P.O. Box 35(YFL) FI-40014 University of Jyvaskyla, Finland}
\author{P.C.~Srivastava}
\affiliation{Department of Physics, Indian Institute of Technology Roorkee, Roorkee 247667, India}
\author{J.~Suhonen}
\affiliation{University of Jyvaskyla, Department of Physics, Accelerator Laboratory, P.O. Box 35(YFL) FI-40014 University of Jyvaskyla, Finland}
\affiliation{International Centre for Advanced Training and Research in Physics (CIFRA), P.O. Box MG12, 077125 Bucharest-M\u{a}gurele, Romania}
\author{M.~Vilen}
\affiliation{University of Jyvaskyla, Department of Physics, Accelerator Laboratory, P.O. Box 35(YFL) FI-40014 University of Jyvaskyla, Finland}
\author{V.~Virtanen}
\affiliation{University of Jyvaskyla, Department of Physics, Accelerator Laboratory, P.O. Box 35(YFL) FI-40014 University of Jyvaskyla, Finland}
\author{A.~Zadvornaya}
\affiliation{University of Jyvaskyla, Department of Physics, Accelerator Laboratory, P.O. Box 35(YFL) FI-40014 University of Jyvaskyla, Finland} 

\date{\today}

\begin{abstract}
The masses of $^{84}$Br, $^{105}$Mo, $^{115,119,121}$Pd, $^{122}$Ag, $^{127,129}$In, $^{132}$Sb and their respective isomeric states have been measured with the JYFLTRAP Penning trap mass spectrometer using the phase-imaging ion-cyclotron-resonance technique. The excitation energies of the isomeric states in $^{132}$Sb and $^{119}$Pd were experimentally determined for the first time, while for $^{84}$Br, $^{115}$Pd and $^{127,129}$In, the precision of the mass values was substantially improved. In $^{105}$Mo and $^{121}$Pd there were no signs of a long-lived isomeric state. The ground-state measurements of $^{119}$Pd and $^{122}$Ag indicated that both are significantly more bound than the literature values. For $^{122}$Ag, there was no indication of a proposed third long-lived state. The results for the $N=49$ nucleus $^{84}$Br and isomers close to doubly magic $^{132}$Sn have been compared to the shell-model and the microscopic quasiparticle-phonon model calculations.
\end{abstract}

\maketitle

\section{Introduction}

Nuclear isomers are long-living excited states of nuclei with half-lives ranging from nanoseconds to billions of years \cite{Walker2020}. The first isomeric state was experimentally observed in 1921 in what is currently known as $^{234}$Pa \cite{Walker2020} by Hahn \cite{Hahn1921}. Since then, about 2000 isomers have been discovered \cite{NUBASE20,Garg2023}. There is no clear cut-off half-life value above which a given state is considered to be isomeric, however, 10- \cite{Garg2023} and 100-ns \cite{NUBASE20} lower limits are currently adopted in the literature. There is also no upper half-life limit, with $^{180}$Ta$^{m}$ ($E_{x,{\rm lit.}} = 76.79(55)$~keV \cite{Nesterenko2022}) being observationally stable \cite{NUBASE20}.

Properties of isomeric states are important in many branches of physics. Two $N=Z$ isomers, namely $^{26}$Al$^{m}$ and $^{38}$K$^{m}$, are used for the determination of the $V_{ud}$ element of the Cabibbo–Kobayashi–Maskawa matrix \cite{Hardy2020}. As it was demonstrated in Ref. \cite{Plattner2023}, even a single measurement of a nuclear structure property of $^{26}$Al$^{m}$ has a significant influence on the corrected $\mathcal{F}t$ value. An ultra-low-lying 8.338(24)-eV isomer in $^{229}$Th \cite{Kraemer2023} is proposed to be used in a nuclear clock \cite{Beeks2021}. Isomeric yield ratios are important observables which are used to benchmark and validate theoretical models \cite{Rakopoulos2019,Gao2023}. Due to differences in spin-parities and half-lives, isomers exhibit different decay patterns with respect to the ground state. This results in an influence on astrophysics models \cite{Misch2020,Misch2021,Misch2021a,Hoff2023}, decay heat \cite{Guadilla2019} as well as reactor antineutrino spectra \cite{Guadilla2019a,Guadilla2022}. It also enables studies of a spin dependence on an exotic $\beta$-delayed fission process \cite{Andel2020}. Because of the energy difference, additional decay channels might be open for isomeric states, for instance a proton decay of $^{53}$Co$^{m}$ \cite{Sarmiento2023}. For some cases, the structure of the isomer and the ground state might differ dramatically, resulting in shape coexistence, as it was observed for $^{79}$Zn \cite{Yang2016,Nies2023}, $^{98}$Y \cite{Cheal2007}, $^{185}$Hg \cite{Marsh2018}, $^{178,187}$Au \cite{Barzakh2020,Cubiss2023} and $^{188}$Bi \cite{Barzakh2021}.  

The most accurate method to extract isomer excitation energies comes from $\gamma$ spectroscopy studies \cite{NUBASE20}. However, for many isomeric states this method is hindered as not all the isomers decay via an internal transition decay. In addition, experimental $\gamma$-spectroscopy techniques might favor production of a single state. For these cases, mass measurements are a good alternative to determine excitation energies. There are several mass measurement techniques available for radioactive nuclei (see e.g. Refs.~\cite{Eronen2016,Yamaguchi2021,Huang2021}) but the resolving power and accuracy they are able to reach can differ by orders of magnitude. To resolve low-lying isomers and to measure reliably the isomer excitation energy, both high resolving power and accuracy are necessary. The experimental technique which fulfils these criteria is Penning trap mass spectrometry \cite{Dilling2018}. Thanks to the recent developments of the phase-imaging ion-cyclotron-resonance technique (PI-ICR) technique \cite{Eliseev2013,Eliseev2014}, the precision is comparable with $\gamma$ spectroscopy. More importantly, this method allows to resolve states lying as close as 10 keV from each other \cite{Huang2021,Orford2018}. 

Mass measurements of isomeric states have been already performed at the Ion Guide Isotope Separator On-Line (IGISOL) facility \cite{Moore2013,Penttila2020} using the JYFLTRAP double Penning trap \cite{Eronen2012}. The previously used time-of-flight ion-cyclotron-resonance technique (TOF-ICR) \cite{Konig1995,Graff1980} enabled studies in Co \cite{Canete2020}, Cu \cite{Canete2024}, Zn \cite{Nies2023}, Y \cite{Hager2007a,Urban2017}, Nb \cite{Hager2007a} and Cd \cite{Kankainen2013} isotopic chains. However, the resolving power was too low to measure isomers in Ru and Rh \cite{Hager2007}. With the commissioning of the PI-ICR method at JYFLTRAP \cite{Nesterenko2018,Nesterenko2021}, we were able to remeasure Rh and Ru isotopic chains \cite{Hukkanen2023,Hukkanen2023a} as well as to extend our studies to other elements: Ag \cite{deGroote2023,Ge2024}, In \cite{Nesterenko2020,Ruotsalainen2023,Jaries2023} and I \cite{Beliuskina2024}. Nevertheless, there are still several isomers whose excitation energy remains unknown or its precision is very low. 

In this work we are reporting the mass measurement of the species with isomeric states produced in proton-induced fission of $^{232}$Th and $^{nat}$U by using the JYFLTRAP double Penning trap. The relevance of the reported results is discussed from the nuclear structure point of view.

\section{Experimental methods}

The experiments were performed at the IGISOL facility \cite{Moore2013,Penttila2020} in Jyv\"askyl\"a, Finland. The radioactive ion beam was produced in fission of either $^{232}$Th or $^{nat}$U targets, both 15~mg/cm$^2$ thick, induced by 25-MeV protons. The thorium target was used to produce $^{84}$Br, $^{105}$Mo and $^{119}$Pd isotopes while the remaining species were produced using the uranium target. The fission products were stopped in a helium-filled gas cell operating at a pressure of about 300 mbars. From there, they were transported by a gas flow into a sextupole ion guide \cite{Karvonen2008} and accelerated by a 30-kV potential. Later, the beam was purified with respect to its mass-over-charge ratio by a 55$^{\circ}$ dipole magnet and it was injected into the buffer gas-filled radio-frequency quadrupole cooler-buncher~\cite{Nieminen2001}. Finally, the cooled and bunched beam was sent to the JYFLTRAP double Penning trap mass spectrometer \cite{Eronen2012}. 

In the first trap, the ions were cooled and mass-selected by means of the buffer-gas cooling technique \cite{Savard1991}. In all the cases except for the $^{127}$In ground state the purified beam was transferred to the second trap and, after a few ms of waiting, it was sent back to the first trap to undergo a second cooling. This additional step in the process enhances the cooling of the ion of interest, thus, allows for a measurement with a better precision. Finally, a clean and cooled bunch of singly-charged ions was injected into the second trap. In there, in a homogeneous magnetic field ${B = 7}$~T \cite{Eronen2012} the ion's cyclotron frequency ${\nu_c = qB/(2 \pi m)}$, which depends on the ion's charge-over-mass ratio $q/m$, was determined using the PI-ICR technique \cite{Eliseev2013,Eliseev2014,Nesterenko2018,Nesterenko2021}.

\begin{figure}
    \centering
    \includegraphics[width=\columnwidth]{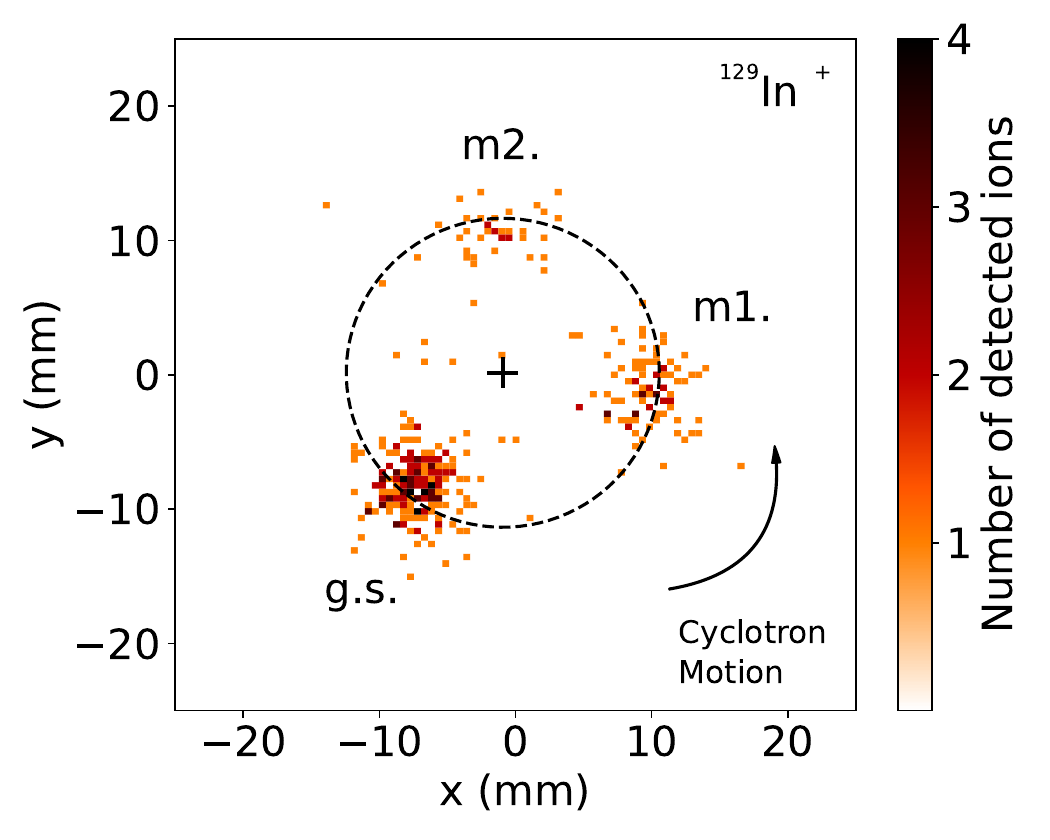}
    \caption{\label{fig:129In_cyc}Projection of the cyclotron motion of $^{129}$In ions onto the position-sensitive detector obtained with the PI-ICR technique using a phase accumulation time $t_{\rm acc} = 208$~ms. The average excitation radius is indicated with the dashed circle and the position of the center spot with the $+$ symbol.}
\end{figure}

In the PI-ICR method, the cyclotron frequency is obtained from phase differences between the ion's cyclotron and magnetron in-trap motions acquired during a phase accumulation time $t_{\rm acc}$ (see Fig.~\ref{fig:129In_cyc}). The $t_{\rm acc}$ value was chosen to avoid an overlap between the cyclotron motion projections of an ion of interest and possible isobaric contaminants, such as isomeric states or molecules. In addition to the accumulation times used for measurements, spectra with few other $t_{\rm acc}$ values were recorded to unambiguously assign the observed states. To determine precisely the magnetic field strength $B$, a cyclotron frequency of reference ions $\nu_{c, \rm ref}$ was measured. During the reported experiments, the reference ions were either surface-ionized $^{133}$Cs (${\mathrm{ME}_{\rm lit.} = -88070.943(8)}$~keV \cite{AME20}) and $^{85}$Rb (${\mathrm{ME}_{\rm lit.} = -82167.341(5)}$~keV \cite{AME20}), delivered from the IGISOL offline surface ion source \cite{Vilen2020} as singly-charged ions, or isobaric species produced together with the ion of interest.

The atomic mass $M$ is connected to the frequency ratio $r=\nu_{c, \rm ref}/\nu_{c}$ between the singly-charged reference ions and the ions of interest:
\begin{equation} \label{eq:mass}
M = r (M_{\rm ref} - m_{e}) + m_{e}\mathrm{,}
\end{equation}
where $M_{\rm ref}$ and $m_e$ are the atomic mass of the reference and the mass of a free electron, respectively. For the cases where the reference ion was an isobaric species, the energy difference between them, $Q$, was extracted as follows:
\begin{equation}
Q = (r-1)[M_{\rm ref} - m_e]c^2 \mathrm{,}
\end{equation} 
with $c$ being the speed of light in vacuum. 

To account for the temporal fluctuations of the magnetic field, the measurements of the ion of interest and the reference ion were alternated. The contribution from electron binding energies have been neglected as it is of the order of a few eV. To account for the ion-ion interaction, a count-rate class analysis was performed \cite{Kellerbauer2003,Roux2013,Nesterenko2021}. For cases where it was not statistically feasible, namely $^{105}$Mo and $^{129}$In, the count rate was limited to one detected ion per bunch. The systematic uncertainties due to the magnetron phase advancement and the angle error, as well as the temporal magnetic field fluctuation of ${\delta B/B = 2.01(25) \times 10^{-12}\mathrm{~min}^{-1}\times \delta t}$, where $\delta t$ is the time between the measurements, were taken into account \cite{Nesterenko2021}. For cases measured against the $^{133}$Cs$^+$ or $^{85}$Rb$^+$ reference ions, a mass-dependent uncertainty of ${\delta r/r = -2.35(81) \times 10^{-10} / \textnormal{u} \times (M_{\rm ref} - M)}$ and a residual systematic uncertainty of ${\delta r/r=9\times 10^{-9}}$ were also added \cite{Nesterenko2021}. 

\section{Results and discussion}

The summary of the measured states and a comparison with the literature \cite{AME20,NUBASE20} is presented in Table \ref{tab:results}. In the following subsections, each case is discussed in detail.

\begin{table*}
\caption{\label{tab:results} The frequency ratios ($r=\nu_{c, \rm ref}/\nu_{c}$), corresponding mass-excess values (ME) and isomer excitation energies ($E_x$) measured in this work using the listed reference ions (Ref.) and a selected accumulation time ($t_{\rm acc}$). The literature mass-excess values (ME$_{\rm lit.}$) and isomer excitation energies ($E_{x, {\rm lit.}}$) are from AME20/NUBASE20~\cite{AME20,NUBASE20}. The differences ${\mathrm{Diff.} = \mathrm{ME}-\mathrm{ME}_{\rm lit.}}$ are provided for comparison. All of the spin-parity assignments $J^{\pi}$ and half-lives $T_{1/2}$ are taken from the NUBASE20 evaluation~\cite{NUBASE20} with an exception of $^{119}$Pd$^{gs,m}$ for which the half-lives are taken from Ref. \cite{Kurpeta2022}. The $(3^+)$ state in $^{122}$Ag is deemed to be non-existent, see text for details. Parentheses indicate a tentative spin-parity assignment whilst \#~denotes values based on systematics.}
\begin{ruledtabular}
\begin{tabular}{lllllllllll}
Nuclide & $J^{\pi}$ & $T_{1/2}$ & Ref. & $t_{\rm acc}$ & $r=\nu_{c, \rm ref}/\nu_{c}$ & ME & ME$_{\rm lit.}$ & $E_x$ & $E_{x, {\rm lit.}}$ & Diff.\\ 
 & &  &  & (ms) &  & (keV) & (keV) &  (keV)  & (keV) & (keV)\\ \hline
$^{84}$Br           & $2^-$     & 31.76(8) m    & $^{85}$Rb     & 380 & \num{0.988278629(20)} & \num{-77767.1(16)} & \num{-77783(40)}  & & & 16(40)\\
$^{84}$Br$^{m}$     & $(6)^-$   & 6.0(2) m      & $^{84}$Br     & 380  & \num{1.000002476(19)} & \num{-77573.5(22)} & \num{-77470(100)}  & \num{193.6(15)} & 310(100) & $-104(100)$\\
$^{105}$Mo          & $(5/2^-)$ & 36.3(8) s     & $^{133}$Cs    & 500 & \num{0.789409786(14)} & \num{-77332.8(21)} & \num{-77331(9)}   & & & $-2(9)$ \\
$^{115}$Pd          & $(1/2)^+$ & 25(2) s       & $^{133}$Cs    & 782.2  & \num{0.864626613(14)} & \num{-80420.5(18)} & \num{-80426(14)}  &  & & 6(14)\\
$^{115}$Pd$^{m}$    & $(7/2^-)$ & 50(3) s       & $^{133}$Cs    & 782.2 & \num{0.864627314(19)} & \num{-80333.7(23)} & \num{-80337(14)}  & 86.8(29) & 89.21(16) & 3(14) \\
$^{119}$Pd          & $3/2^+\#$ & 0.88(2) s     & $^{119}$Pd$^{m}$    & 284 & \num{0.999998203(27)}& \num{-71609.7(42)} & \num{-71407(8)}  &  & & $-203(9)$\\
$^{119}$Pd$^{m}$    & $11/2^-\#$& 0.85(1) s     & $^{133}$Cs    & 284 & \num{0.894796098(23)} & \num{-71410.6(29)} & \num{-71110(150)}\#  & 199.1(30) & 300(150)\# & $-301(150)$\# \\
$^{121}$Pd          & $3/2^+\#$ & 290(1) ms     & $^{133}$Cs    & 329 & \num{0.909886693(30)} & \num{-66181.4(37)} & \num{-66182(3)}  &  & & 1(5) \\
$^{122}$Ag          & $(3^+)$  	& 529(13) ms    & -    & - & -    	& non-existent 	& \num{-71110(40)}      &              & & \\
$^{122}$Ag$^{m1}$          & $(1^-)$  	& 550(50) ms    & $^{133}$Cs    & 200 & \num{0.917370167(42)}    	& \num{-71220.2(52)} 	& \num{-71030(60)}\#      & 0             & 80(50)\# & $-190(60)$\# \\
$^{122}$Ag$^{m2}$    & $(9^-)$   & 200(50) ms    & $^{122}$Ag    & 200 & \num{1.000 002 674(44)}    	& \num{-70916.5(72)} 	& \num{-71030(60)}\#  & 303.7(50)   & 80(50)\# & $114(60)$\#\\
$^{127}$In          & $9/2^+$ & 1.086(7) s    & $^{133}$Cs    & 450 & \num{0.954945243(10)} & \num{-76891.2(13)} & \num{-76880(10)}  & & & $-11(10)$\\
$^{127}$In$^{m1}$   & $1/2^-\#$ & 3.618(21) s   & $^{127}$In    & 450 & \num{1.000003450(42)} & \num{-76483.3(52)} & \num{-76486(15)}  & \num{407.9(50)} & 394(18)\footnotemark[1] &  3(16)\\
$^{127}$In$^{m2}$   & $(21/2^-)$& 1.04(10) s    & $^{127}$In    & 450 & \num{1.000014622(10)} & \num{-75162.5(18)} & \num{-75110(40)}\footnotemark[2]  & 1728.7(12) & 1770(40)\footnotemark[3] & $-53(40)$\\
$^{129}$In$^{m1}$   & $1/2^-$ & 1.23(3) s   & $^{129}$In    & 208 & \num{1,000003739(50)} & \num{-72385.8(63)} & \num{-72384.2(20)}   & 449.1(59) & 450.73(16) & $-2(7)$\\
$^{129}$In$^{m2}$   & $(23/2^-)$& 670(100) ms   & $^{129}$In    & 208 & \num{1.000013711(28)} & \num{-71188.3(39)} & \num{-71180(50)}  & 1646.6(33) & 1650(50) & $-8(50)$ \\
$^{132}$Sb          & $(4)^+$   & 2.79(7) m     & $^{133}$Cs    & 471 & \num{0.992544008(21)} & \num{-79629.6(27)} & \num{-79635.3(25)}  & & & 6(4)\\
$^{132}$Sb$^{m}$    & $(8^-)$   & 4.10(5) m     & $^{132}$Sb    & 471 & \num{1.000001134(16)} & \num{-79490.3(33)} & \num{-79490(50)}\#  & 139.3(20) & 150(50)\#\footnotemark[4] & 0(50)\#\\
\end{tabular}
\footnotetext[1]{Also 408.0(3)~keV in Ref. \cite{Lorenz2019} and 406(12)~keV in Ref. \cite{Izzo2021}.}
\footnotetext[2]{Also \num{-75126(36)}~keV in Ref. \cite{Izzo2021}.}
\footnotetext[3]{Also 1744(9)~keV in Ref. \cite{Izzo2021}.}
\footnotetext[4]{Also 153(14)~keV in Ref. \cite{Kankainen2013}.}
\end{ruledtabular}
\end{table*}

\subsection{$^{84}$Br}

All of the $N=49$ isotones lying between $Z = 28$ and $Z = 50$ have at least one isomeric state \cite{NUBASE20,Pedersen2023}. In $^{84}$Br, two long-lived states are known. The mass of the ground state is deduced from two $\beta$-decay studies, $^{84}$Se to $^{84}$Br \cite{Rengan1968} and $^{84}$Br to $^{84}$Kr \cite{Hattula1970}, while in case of the isomer, the mass is deduced from a single $\beta$-decay measurement of $^{84}$Br$^{m}$ to $^{84}$Kr \cite{Hattula1970}.

In this work, the ground-state mass was measured against $^{85}$Rb with 380~ms accumulation time. The extracted mass-excess value, $\mathrm{ME} = -77767.1(16)$~keV, is in agreement with AME20 ($\mathrm{ME}_{\rm lit.} = -77783(40)$~keV \cite{AME20}) but it is 25 times more precise. 

To determine the isomer excitation energy, three different accumulation times, 250, 260 and 380~ms were tested and the final measurement was performed with $t_{\rm acc} = 380$~ms. The extracted value, ${E_x = 193.6(15)}$~keV, is 116(100)~keV lower than the literature (${E_{x,{\rm lit.}} = 310(100)}$~keV \cite{NUBASE20}) and it is 67 times more precise.

\begin{figure}
    \centering
    \includegraphics[width=\columnwidth]{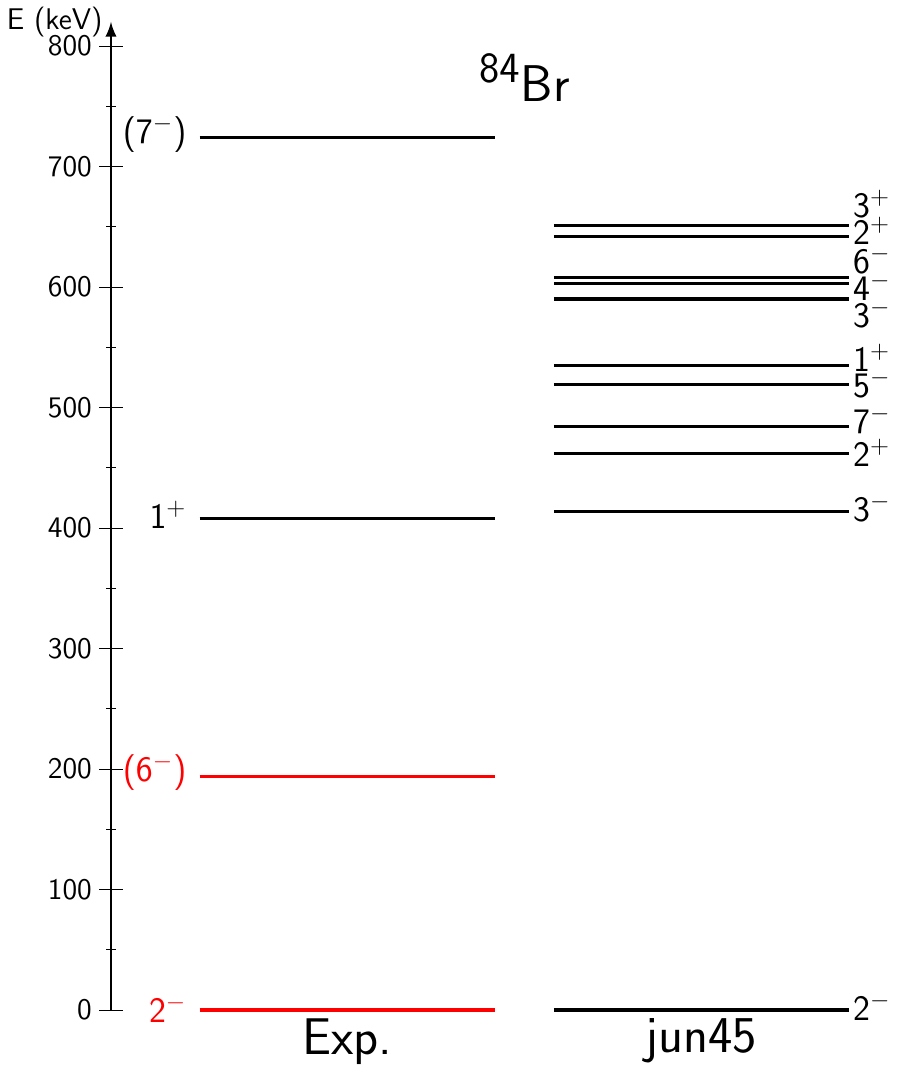}
    \caption{\label{fig:84BrSM}Comparison experimental and theoretical excited states in $^{84}$Br up to 800 keV. The experimental data is adapted from Ref. \cite{ENSDF} based on the isomer excitation energy measured in this work, indicated in red.}
\end{figure}

In order to better understand the underlying structure, shell model calculations with the jun45 interaction \cite{Honma2009} using {\sc kshell} code \cite{Shimizu2013,Shimizu2019} were performed. The comparison between the theoretical and experimental results is presented in Fig.~\ref{fig:84BrSM}. The calculated negative-parity states do not reproduce the experimental data. While the predicted $2^-$ ground state is in agreement with the literature \cite{Hattula1970}, the properties of the isomer are not reproduced. The theory suggests that it is a $7^-$ state and its excitation energy is overestimated by about 300 keV. The energy of the only known low-lying positive-parity state, the $1^+_1$ state, is reproduced relatively well. 

The dominant configuration of the $2^-$ ground state and the $7^-$ state in the shell model calculations consists of an unpaired proton on the $f_{5/2}$ shell and an unpaired neutron on the $g_{9/2}$ shell. However, for the $6^-$ state, the main configuration is $\pi p_{3/2} \otimes \nu g_{9/2}$. Since the isomeric spin-parity is tentative and based on a single $\beta$-decay study \cite{Hattula1970}, further studies, for instance a laser spectroscopy experiment, are needed to determine the nuclear spin and the magnetic moment which can validate our calculations. 

\subsection{$^{105}$Mo}

The decay of $^{105}$Mo has been studied in several experiments \cite{Wilhelmy1970,Kaffrell1976,Kiso1977,Tittel1977} but there are discrepancies in the determined half-lives. In particular, in Ref.~\cite{Kiso1977}, where the $^{105}$Mo nuclei were produced in thermal neutron-induced fission of $^{235}$U, two groups of half-lives, around 30 and 50~s, were obtained. A possible explanation might be the presence of an isomer. However, in the TOF-ICR mass measurement performed at IGISOL, during which the radioactivity was produced with 25-MeV proton-induced fission of $^{nat}$U, only one state was observed  \cite{Hager2006}. With the limited resolving power of the TOF-ICR technique used in Ref. \cite{Hager2006}, it would not have been possible to resolve two states with an energy difference below 400~keV. While only one state was present in the laser spectroscopy experiment performed at IGISOL with radioactivity produced in 33-MeV proton-induced fission of $^{nat}$U \cite{Charlwood2009}, the authors claim that the statistics obtained for this case was limited. 

\begin{figure}
    \centering
    \includegraphics[width=\columnwidth]{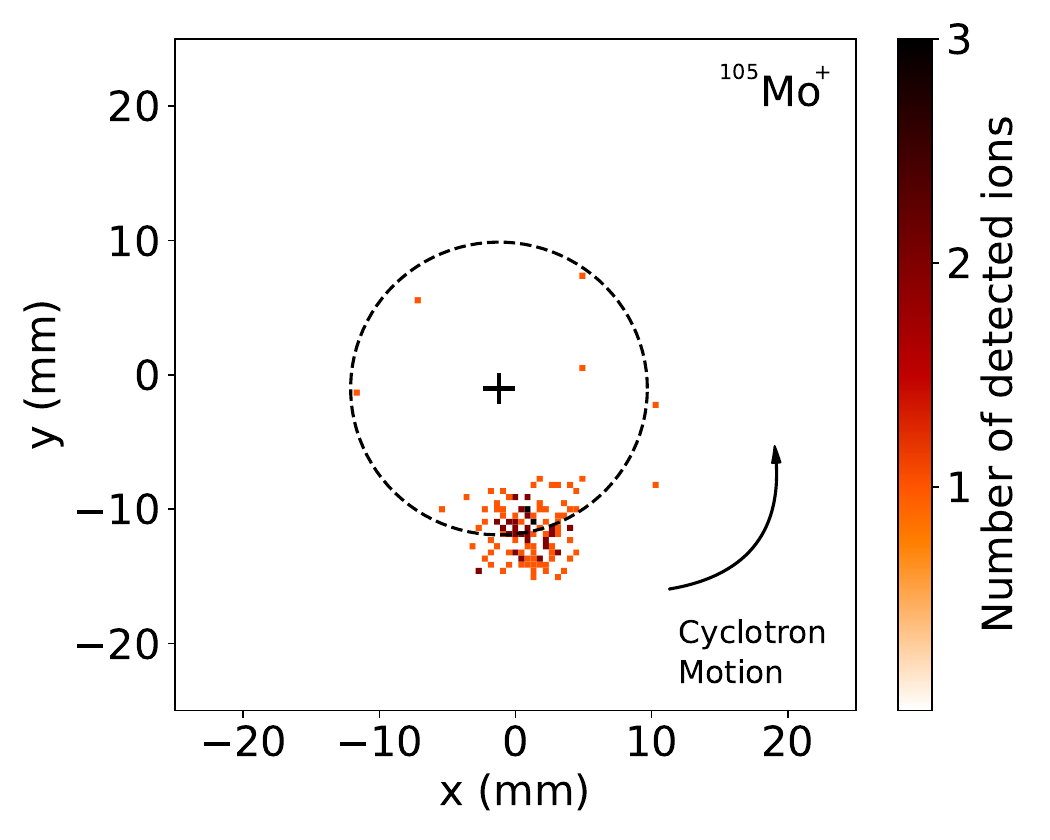}
    \caption{\label{fig:105Mo_cyc}Projection of the cyclotron motion of $^{105}$Mo ions onto the position-sensitive detector obtained with the PI-ICR technique using a phase accumulation time $t_{\rm acc} = 500$~ms. The average excitation radius is indicated with the dashed circle and the position of the center spot with the $+$ symbol. Only one state is present.}
\end{figure}

In this work we searched for a possible new isomeric state using the PI-ICR technique. Only one state was observed, see Fig.~\ref{fig:105Mo_cyc}, and its mass was measured against $^{133}$Cs with a 500-ms accumulation time. The extracted mass-excess value, $\mathrm{ME} = -77332.8(21)$~keV, is in agreement with the AME20 value ($\mathrm{ME}_{\rm lit.} = -77331(9)$~keV \cite{AME20}) which is based mostly on the previous JYFLTRAP measurement \cite{Hager2006}. However, our refined result is four times more precise. Considering that the mass of $^{105}$Mo from this work and from Ref. \cite{Hager2006} differ by only 2(9)~keV despite different target material used ($^{232}$Th in this work, $^{nat}$U in Ref. \cite{Hager2006}), it is unlikely an isomer exists. At the same time, if it exists and its production rate is at least 10\% of the one of the ground state, the upper limit for the excitation energy can be estimated to be 25~keV based on the experimental conditions and the utilized accumulation time.

\subsection{$^{115,119,121}$Pd}

There are two known long-lived states in $^{115}$Pd and the isomer excitation energy, ${E_{x,{\rm lit.}} = 89.21(16)}$~keV, is known precisely from $\gamma$-spectroscopy studies \cite{ENSDF,NUBASE20}. The masses of both states have been previously measured at JYFLTRAP using the TOF-ICR method \cite{Hager2007}. However, due to the low excitation energy of the isomer, which was at the edge of the resolving power, an additional uncertainty was added to the final result \cite{Hager2007}. In this work, both states were measured using the PI-ICR method against $^{133}$Cs. The extracted mass-excess values, ${\mathrm{ME}(^{115}\mathrm{Pd}^{gs}) = -80420.5(18)}$~keV and ${\mathrm{ME}(^{115}\mathrm{Pd}^{m}) = -80333.7(23)}$~keV, agree with the previous measurements but they are eight and six times more precise, respectively. The extracted isomer excitation energy, $E_{x} = 86.8(29)$~keV, also agrees with the literature \cite{NUBASE20}. 

An isomeric state in $^{119}$Pd was observed for the first time in the $\beta$-decay study performed at IGISOL \cite{Kurpeta2022}. In addition to the $\gamma$ rays associated with the $\beta$ decay of two states in $^{119}$Pd, $\gamma$-ray transitions uncorrelated with $\beta$ particles were observed. Their presence was proposed to be associated with an internal transition decay of $^{119}$Pd$^{m}$ and it might suggest that the isomer excitation energy is at least 240~keV \cite{Kurpeta2022}. 

\begin{figure}
    \centering
    \includegraphics[width=\columnwidth]{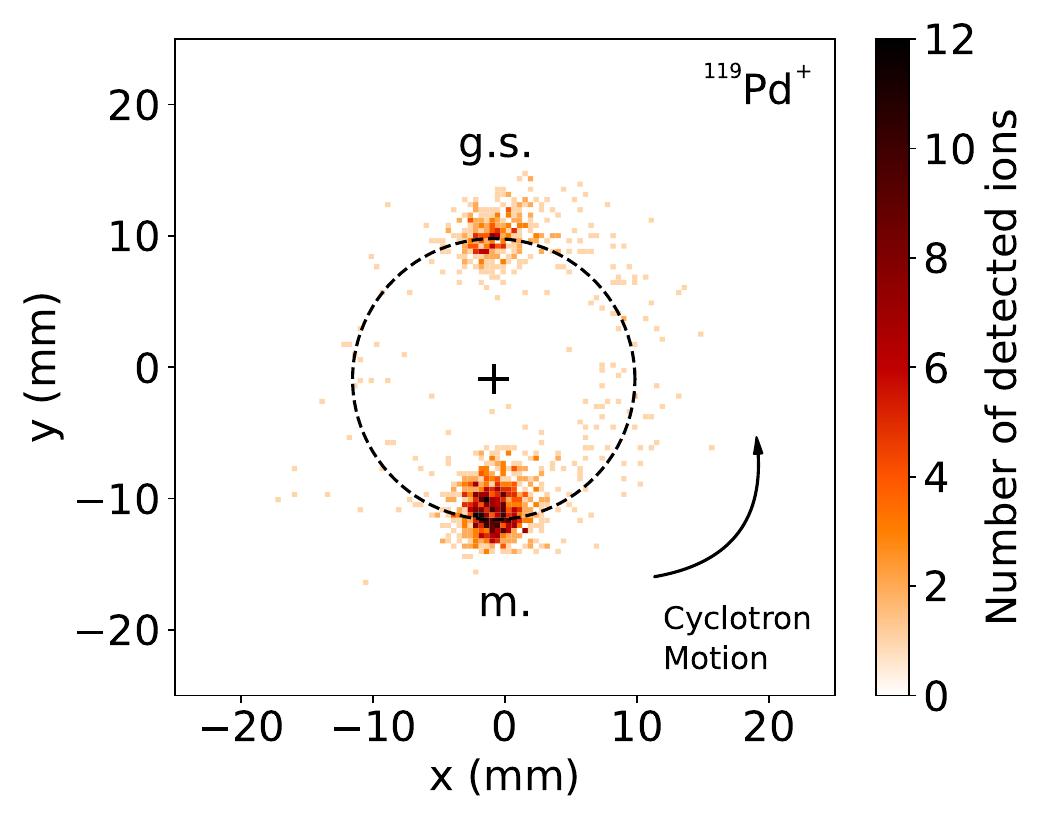}
    \caption{\label{fig:119Pd_cyc}Projection of the cyclotron motion of $^{119}$Pd ions onto the position-sensitive detector obtained with the PI-ICR technique using a phase accumulation time $t_{\rm acc} = 284$~ms. The average excitation radius is indicated with the dashed circle and the position of the center spot with the $+$ symbol.}
\end{figure}

In this work, two states were observed in $^{119}$Pd (see Fig. \ref{fig:119Pd_cyc}). The isomeric state was measured against $^{133}$Cs$^+$ ions with a 284-ms accumulation time. The extracted mass, $\mathrm{ME} = -71410.6(29)$~keV, differs from the NUBASE20 extrapolation (${\mathrm{ME}_{\rm lit.} = -71110(150)\#}$~keV \cite{NUBASE20}) by 301(150)\#~keV. The isomer excitation energy was determined to be $E_x = 199.1(30)$~keV. It is in agreement with the NUBASE20 extrapolation (${E_{x,{\rm lit.}} = 300(150)\#}$~keV \cite{NUBASE20}) but below the lower limit of 240~keV proposed in Ref.~\cite{Kurpeta2022}.

Our new result yields a ground-state mass excess of ${\mathrm{ME} = -71609.7(42)}$~keV for $^{119}$Pd. It is thus found to be 203(9)~keV more bound than in literature (${\mathrm{ME}_{\rm lit.} = -71407(8)}$~keV \cite{AME20}). We note that the currently known ground-state mass of $^{119}$Pd \cite{Hager2007,AME20} corresponds exactly to the mass of the isomeric state reported in this work. This fact suggests that in the previous study \cite{Hager2007} only the isomeric state was measured.  

\begin{figure}
    \centering
    \includegraphics[width=\columnwidth]{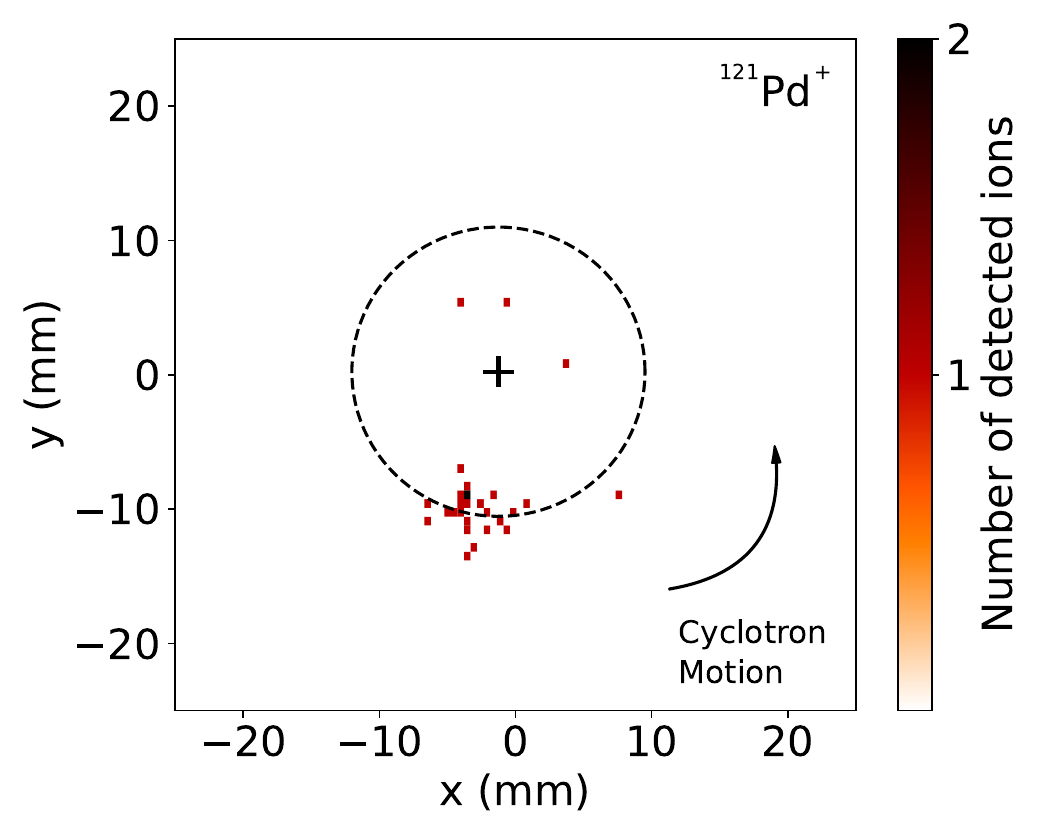}
    \caption{\label{fig:121Pd_cyc}Projection of the cyclotron motion of $^{121}$Pd ions onto the position-sensitive detector obtained with the PI-ICR technique using a phase accumulation time $t_{\rm acc} = 329$~ms. The average excitation radius is indicated with the dashed circle and the position of the center spot with the $+$ symbol.}
\end{figure}

The mass of $^{121}$Pd was measured previously at JYFLTRAP using the TOF-ICR method \cite{Hakala2011}. However, considering the presence of long-lived isomers in $^{115,117,119}$Pd \cite{NUBASE20}, as well as the proposed long-lived $\beta$-decaying isomers in $^{123,125}$Pd \cite{Chen2019,NUBASE20}, we remeasured $^{121}$Pd using the PI-ICR technique as it provides better resolving power. Only one state was observed in this work, see Fig.~\ref{fig:121Pd_cyc}. The measured mass-excess value, ${\mathrm{ME} = -66181.4(37)}$~keV, agrees with the literature (${\mathrm{ME}_{\rm lit.} = -66182(3)}$~keV \cite{NUBASE20}). If the isomeric state exists, has a half-life of the same order of magnitude as the ground state and a fission yield is at least 40\% of the one of the ground-state, its excitation energy would be below 35~keV. This estimation is made considering the experimental conditions and the used accumulation time. We note that if the half-life of the hypothetical isomer is shorter, the production yield needs to be larger to compensate for the decay losses.


\begin{figure}
\centering
\includegraphics[width=\columnwidth]{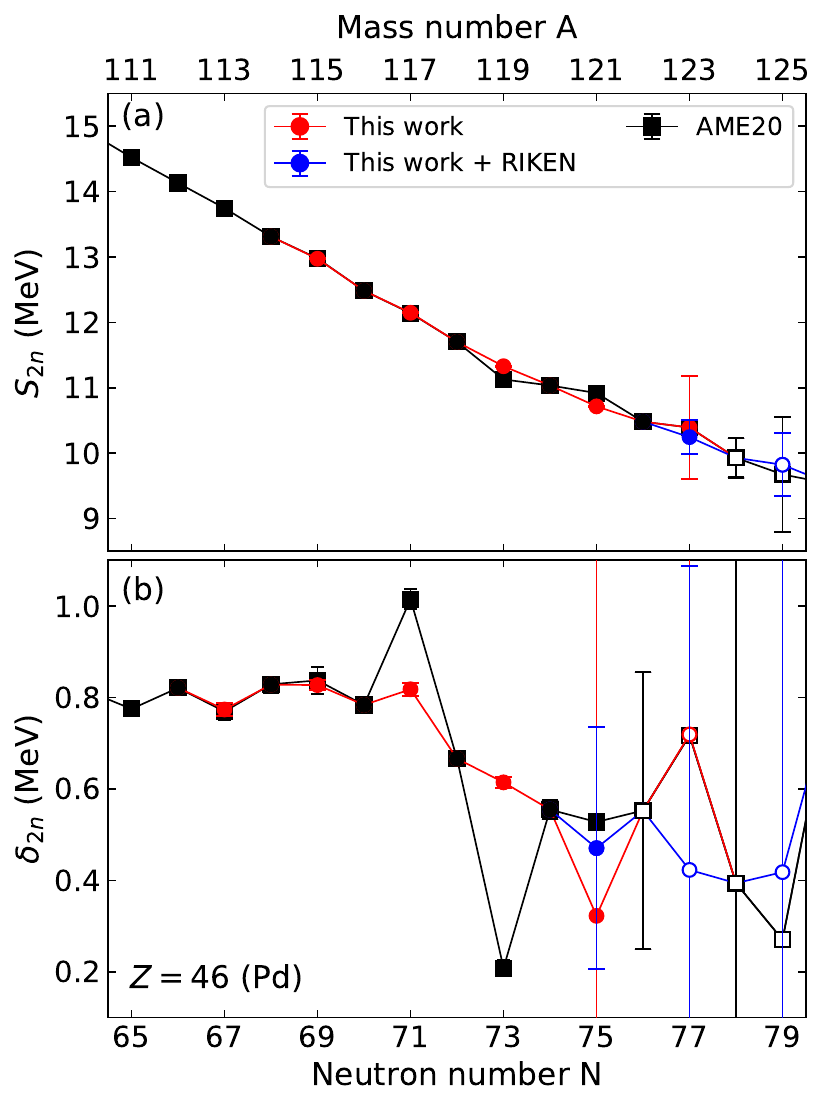}
\caption{\label{fig:Pdsystematics}A comparison of a) two-neutron separation energies $S_{2n}$ and b) two-neutron shell-gap energies $\delta_{2n}$ for the Pd isotopic chain between AME20 \cite{AME20} and the results from this work. In addition, a mass value for $^{123}$Pd measured at RIKEN \cite{Li2022} was added for comparison. The empty symbols indicate that the value is partially or fully based on the AME20 extrapolations.}
\end{figure}

To analyze the influence of the new mass values on the mass trends in palladium, we calculated the two-neutron separation energies $S_{2n}$:
\begin{equation}
S_{2n}(Z,N) = \mathrm{ME}(Z,N-2) - \mathrm{ME}(Z,N) +2\mathrm{ME}_n \mathrm{,}
\end{equation}
and the two-neutron shell-gap energies $\delta_{2n}$:
\begin{equation}
\delta_{2n}(Z,N) = S_{2n}(Z,N) - S_{2n}(Z,N+2) \mathrm{,}
\end{equation}
where $\mathrm{ME}(Z,N)$ is a mass excess of a nucleus with given proton ($Z$) and neutron ($N$) numbers and $\mathrm{ME}_n$ is the mass excess of a free neutron. The mass trends can reveal information regarding changes in nuclear structure of the ground state, such as shell closures or an onset of deformation \cite{Lunney2003,Eronen2016,Garrett2022}. The comparison of results from this work and AME20 is presented in Fig.~\ref{fig:Pdsystematics}.

A significant staggering of the $S_{2n}$ values around ${N=73}$ and ${N=75}$ disappears completely when the refined $^{119}$Pd ground-state mass value is included, see Fig.~\ref{fig:Pdsystematics}a. This effect is better visible when analysing the $\delta_{2n}$ curve, see Fig.~\ref{fig:Pdsystematics}b. With the new mass value the trend becomes more linear and two strong peaks at ${N=71}$ and ${N=73}$ disappear. These results indicate an absence of significant structural changes around ${N=73}$. We note that a peak in the $S_{2n}$ curve at $N=77$ and a drop in the $\delta_{2n}$ curve at ${N=75}$ disappear when the updated mass value of $^{123}$Pd from Ref.~\cite{Li2022} is used. 

\subsection{$^{122}$Ag}

\begin{figure}[h!t!b]
    \centering
    \includegraphics[width=\columnwidth]{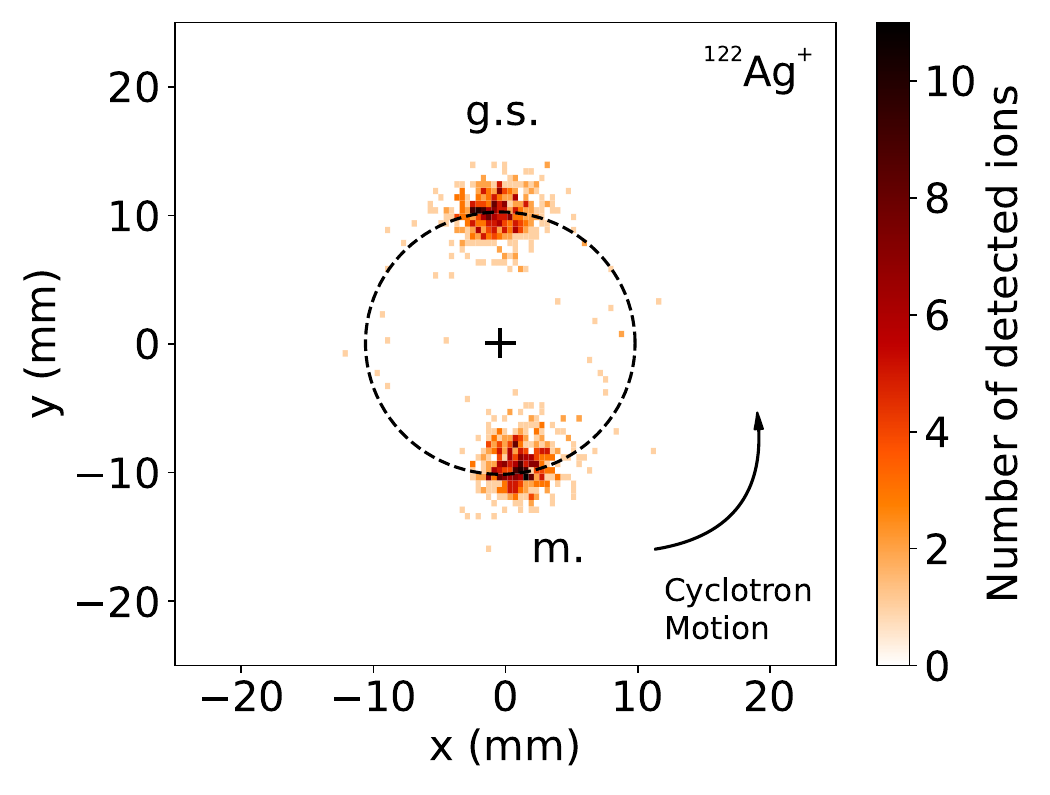}
    \caption{\label{fig:122Ag_cyc}Projection of the cyclotron motion of $^{122}$Ag$^+$ ions onto the position-sensitive detector obtained with the PI-ICR technique using a phase accumulation time $t_{\rm acc} = 200$~ms. The average excitation radius is indicated with the dashed circle and the position of the center spot with the $+$ symbol.}
\end{figure}

Three long-lived states in $^{122}$Ag are reported in the NUBASE20 \cite{NUBASE20} and the ENSDF \cite{ENSDF} evaluations. The excitation energies of the two isomers are unknown and the order of all three states is uncertain. In our measurements only two long-lived states were observed, see Fig. \ref{fig:122Ag_cyc}. The phase-accumulation time $t_{\rm acc}$ in the PI-ICR measurement was varied from 50 to 200 ms and, for the final measurement, $t_{\rm acc} = 200$~ms was selected.

The ground state was measured against $^{133}$Cs and its mass value, $\mathrm{ME} = -71220.2(52)$~keV, is 110(40)~keV more bound than in AME20 \cite{AME20}. The isomer excitation energy, $E_x = 303.7(50)$~keV, is 224(50)~keV higher than the NUBASE20 extrapolation ($E_{x,{\rm lit.}} = 80(50)\#$~keV \cite{NUBASE20}). 

It should be noted that the AME20 value is based on the ISOLTRAP measurement reported in Ref. \cite{Breitenfeldt2010}. However, the procedure for unresolved isomeric mixtures \cite{Huang2021} has been applied in AME20 and therefore, the AME20 value differs from the original mass-excess value of ${\mathrm{ME} = -71066(17)}$~keV \cite{Breitenfeldt2010}. Nevertheless, the literature mass value lies in between the masses of the ground state and the isomer. This behavior indicates that an unresolved mixture was measured and it is consistent with other studies of low-lying isomeric states \cite{Hukkanen2023,Hukkanen2023a}.

An apparent absence of the third state can be explained by a small energy difference ($\leq 25$ keV) between either of the two states or a small production cross section in fission ($\leq 3\%$ of the ground-state yield). However, a more likely explanation, as detailed below, is that there are only two long-lived states present in $^{122}$Ag, a low-spin ground state and a high-spin isomer.

According to the NUBASE20 and ENSDF evaluations \cite{NUBASE20,ENSDF}, the three $\beta$-decaying states are tentatively assigned spins and parities $3^+$, $1^-$ and $9^-$. The $3^+$ assignment for the ground state comes from the $\beta$-decay study \cite{Shih1978} where large $\beta$-feedings and low $\log(ft)$ values to the $2^+$ and $4^+$ states in $^{122}$Cd were observed. However, in that work a beam of $^{122}$Ag, which was produced in neutron-induced fission, was not isomerically purified. More recent experimental studies, which have used similar production mechanisms \cite{Zamfir1995,Kratz2000}, indicated a presence of two long-lived states in the beam. Thus, the $3^+$ spin assignment based on the $\log(ft)$ values is not correct.

The analysis of the half-lives further supports the two-state hypothesis. The $3^+$-level half-life of 0.529(13)~s \cite{ENSDF} is an average of two $\beta$-delayed neutron measurements \cite{Reeder1983,Fedoseyev1995}. However, in the in-source laser spectroscopy study performed at ISOLDE \cite{Kratz2000}, two long-lived states in $^{122}$Ag were observed and both of them are $\beta$-delayed-neutron emitters. Therefore, since there was no isomeric purification in Refs. \cite{Reeder1983,Fedoseyev1995}, the half-life has had two components and the ENSDF average is of an unknown mixture. 

There are two more half-life measurements reported in the literature, 0.48(8)~s \cite{Shih1978} and 0.357(24)~s \cite{Montes2006}. Both of them are in between 0.20(5)~s and 0.55(5)~s, reported in the laser spectroscopy study \cite{Kratz2000} for the high- and low-spin states, respectively, as well as in between 0.29(5)~s and 0.8(2)~s, reported in Ref.~\cite{Tomlin2006}. This indicates that the beam contained a mixture of isomers, as explicitly stated in Ref.~\cite{Montes2006}. 

In this work, the difference between the isomeric state and the ground state decay constants ${\Delta \lambda=\lambda_{m}-\lambda_{gs}}$ was extracted by varying waiting times (from 0 to 0.5 s) in the cooler before injecting the beam into the Penning traps. The obtained difference, ${\Delta\lambda = 1.1(3) ~1/\mathrm{s}}$, is in good agreement with ${\Delta\lambda=2.2(9)~1/\mathrm{s}}$ extracted from the half-lives reported in Ref. \cite{Kratz2000} and ${\Delta\lambda=1.5(5)~1/\mathrm{s}}$ extracted from the values in Ref.~\cite{Tomlin2006}.

The isomeric ratio of both spots observed on the PI-ICR plot is close to 1. However, the half-life of the high-spin state is shorter than the low-spin state \cite{Kratz2000} and, in addition, it is shorter than the measurement cycle (${t \approx 470}$~ms). After correcting for the decay losses, the yield of the high-spin state is higher, in line with Ref.~\cite{Rakopoulos2019}.

Considering the presented facts, we assign the ground-state as a low-spin $(1^-)$ state with ${T_{1/2} = 550(50)}$~ms and the isomer as the high-spin $(9^-)$ state with ${T_{1/2} = 200(50)}$~ms. We also deem the $(3^+)$ state with ${T_{1/2} = 529(13)}$~ms to be non-existent.

\subsection{$^{127,129}$In}

$^{127,129}$In isotopes have very high spin isomeric states, $(21/2^-)$ and $(23/2^-)$, respectively. They were observed for the first time at the OSIRIS facility in Studsvik \cite{Gausemel2004} and their excitation energies in the previous editions of AME, $E_{x,{\rm lit.}}(^{127}\mathrm{In}^{m2}) = 1870(60)$~keV and $E_{x,{\rm lit.}}(^{129}\mathrm{In}^{m2}) = 1660(50)$~keV \cite{NUBASE16},  have been established by measuring their total decay energies. Two recent $^{127}$In$^{m2}$ excitation energy measurements performed at TITAN in TRIUMF showed significant deviations from the literature and from each other, with $1697(49)$~keV reported in Ref. \cite{Babcock2018} and $1744(9)$~keV in Ref. \cite{Izzo2021}. In addition, in Ref. \cite{Izzo2021} the excitation energy of $^{129}\mathrm{In}^{m2}$, $1649(82)$~keV, was measured. While it is in agreement with the literature, the precision is limited due to a presence of another long-lived state.

In this work, the ground-state mass of $^{127}$In was measured against $^{133}$Cs while both isomeric states were measured against the ground state. For all cases, the accumulation time was 450~ms. The extracted ground-state mass value, ${\mathrm{ME} = -76891.2(13)}$~keV is 11 keV more bound and eight times more precise than  in AME20 \cite{AME20}. The excitation energy of the first isomeric state, ${E_{x} = 407.9(50)}$~keV, is in agreement with the NUBASE20 value (${E_{x,{\rm lit.}} = 394(18)}$~keV \cite{NUBASE20}) as well as with the results from TITAN ($406(12)$~keV \cite{Izzo2021}) and with the more precise value from the $^{127}$Cd $\beta$-decay spectroscopy study performed at IGISOL ($408.0(3)$~keV \cite{Lorenz2019}).

The excitation energy of the second isomeric state from this work, ${E_{x} = 1728.7(12)}$~keV, is 41(40)~keV lower than the NUBASE20 value \cite{NUBASE20} which is weighted 58\% on the TITAN measurement \cite{Babcock2018} and 42\% on the $\beta$-decay OSIRIS study \cite{Gausemel2004,Huang2021}. It is also 15(9)~keV lower than the more recent TITAN value \cite{Izzo2021}.

The excitation energies of the first and the second isomeric states in $^{129}$In were measured against the ground state with 208~ms accumulation time. The extracted values, ${E_x(^{129}\mathrm{In}^{m1}) = 449.1(59)}$~keV and ${E_x(^{129}\mathrm{In}^{m2}) = 1646.6(33)}$~keV, respectively, are both in agreement with NUBASE20 \cite{NUBASE20} as well as the TITAN measurement \cite{Izzo2021}. In addition, in the case of $^{129}$In$^{m2}$, our result is 15 times more precise. It should be noted that since the masses of the second and the third isomeric states are connected via a 281-keV $\gamma$-ray transition \cite{NUBASE20,Garcia2021}, our refined mass value of $^{129}$In$^{m2}$ leads to an improved precision for the $^{129}$In$^{m3}$ excitation energy, ${E_x(^{129}\mathrm{In}^{m3}) = 1927.6(33)}$~keV. 

The excitation energies of the $1/2^-$ states, the $21/2^-$ state in $^{127}$In and the $23/2^-$ and the $29/2^+$ states in $^{129}$In were calculated within the shell-model framework with the SNET interaction \cite{Hosaka1985,Brown1994} and the {\sc kshell} code. To make the calculations more feasible, the valence space was truncated by allowing only up to two protons onto the $\pi g_{7/2}$ shell and by completely restricting neutron excitations across the $N=50$ magic number. 

In the case of the shell model, the success in predicting the relative order of states in an odd-mass nucleus is based on the original fitting scheme of the adopted single-particle orbitals and the two-body interaction. In the case of the SNET interaction, the single-particle energies (SPE) and two-body matrix elements (TBME) for the proton-proton interaction below $Z=50$ ($\pi 0f_{5/2}$, $\pi 1p_{3/2}$, $\pi 1p_{1/2}$, and $\pi 0g_{9/2}$ orbitals) were obtained from a least-squares fit to energy levels of the $N=50$ isotones \cite{Ji1988}. The neutron-neutron interaction above $N=50$ ($\nu 0g_{7/2}$, $\nu 1d_{5/2}$, $\nu 1d_{3/2}$, $\nu 2s_{1/2}$, and $\nu 0h_{11/2}$ orbitals) was constructed starting with a set of TBME obtained from a similar least-squares fit to the $N=82$ isotones with a $^{132}$Sn core \cite{Kruse1982} in which the protons fill the same set of orbitals as do the neutrons outside of the $^{100}$Sn core. Then, a calculated Coulomb interaction was subtracted and the resulting TBME was scaled by a factor of $(132/100)^{0.3}$. 

\begin{figure}
    \centering
    \includegraphics[width=\columnwidth]{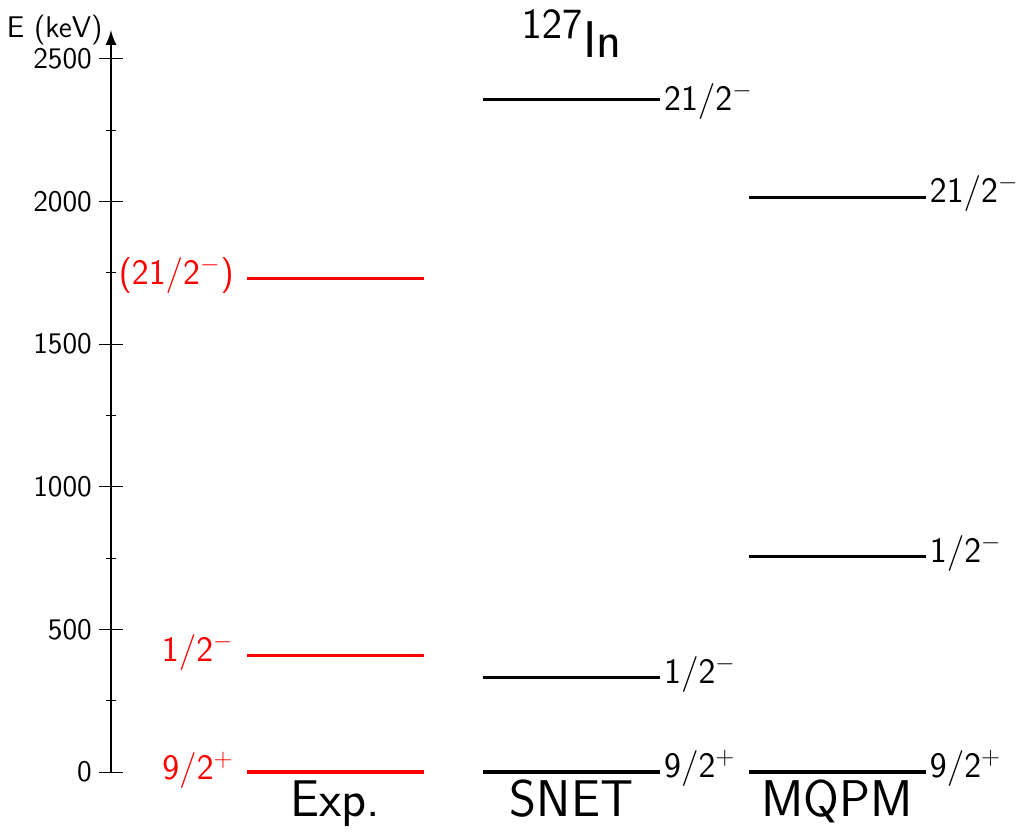}
    \caption{\label{fig:127InSM}Comparison of experimental and theoretical isomeric states in $^{127}$In. States measured in this work are indicated in red.}
\end{figure}

\begin{figure}
    \centering
    \includegraphics[width=\columnwidth]{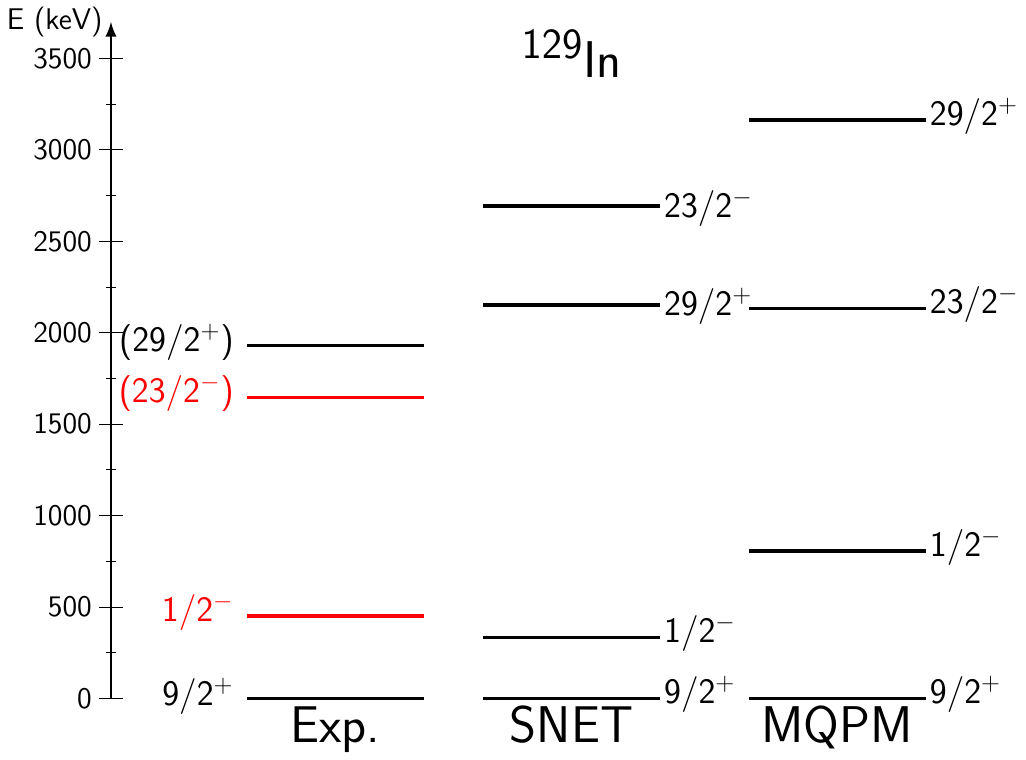}
    \caption{\label{fig:129InSM}Comparison of experimental and theoretical isomeric states in $^{129}$In. The excitation energy of the $(29/2^+)$ isomer is based on the internal transition energy reported in Ref. \cite{Garcia2021}. States measured in this work are indicated in red.}
\end{figure}

The proton-neutron interaction is basically constructed from a G-matrix interaction and expressed as a superposition of one-boson-exchange potential (OBEP) type functions, whose oscillator matrix elements fit the G-matrix elements derived from the Paris potential \cite{Lacombe1980}. The G-matrix elements can explain the low-energy properties (around the Fermi surface), while the behavior of a G-matrix interaction at higher angular momenta fails sometimes, as can be seen in Figs.~\ref{fig:127InSM} and~\ref{fig:129InSM}.

Additional calculations were performed within the microscopic quasiparticle-phonon model (MQPM) framework \cite{Toivanen1998,Ejiri2019}. The MQPM is based on the quasiparticle random-phase approximation (QRPA) for even-even nuclei \cite{Suhonen2007} and an extra proton (neutron) quasiparticle is coupled to the QRPA excitations using the residual Hamiltonian \cite{Toivanen1998} in order to access proton-odd (neutron-odd) nuclei. MQPM has the advantage that when the energy spectrum of the neighboring even-even reference nucleus is well reproduced by adjusting the residual Hamiltonian, then the three-quasiparticle states in the odd-mass nucleus are relatively well reproduced, as in the present case. 

For $^{127}$In, the MQPM states were created by coupling a proton quasiparticle to $^{128}$Sn, which was used as the reference nucleus for which the QRPA calculation was carried out, while for $^{129}$In the neighboring $^{130}$Sn served as the reference nucleus. In the QRPA calculation for $^{128}$Sn, the phonon energies were 0.965 MeV ($2^+_1$), 1.936 MeV ($5^-_1$), 1.830 MeV ($7^-_1$) and 2.746 MeV ($10^+_1$), corresponding to the experimental $^{128}$Sn states 1.169 MeV ($2^+$), 2.121 MeV ($5^-$), 2.092 MeV ($7^-$) and 2.434 MeV ($10^+$), respectively \cite{ENSDF}. For $^{130}$Sn, the phonon energies were 1.255 MeV ($2^+_1$), 2.281 MeV ($5^-_1$), 2.047 MeV ($7^-_1$), and 3.081 MeV ($10^+_1$), corresponding to the experimental $^{130}$Sn states 1.221 MeV ($2^+$), 2.085 MeV ($5^-$), 1.947 MeV ($7^-$) and 2.434 MeV ($10^+$), respectively \cite{ENSDF}. The relatively high energy for the $10^+_1$ phonon in $^{130}$Sn may partially explain why the isomeric state $29/2^+$ is notably above the experimental value, while the other states, also consistently above the corresponding experimental values, are closer to the measured energies.

\subsection{$^{132}$Sb}

The $^{132}$Sb isotope lies in the direct vicinity of the doubly magic $^{132}$Sn nucleus. Despite the two long-lived states being known since 1956 \cite{NUBASE20}, the isomer excitation energy remains unknown. In this work, both long-lived states were measured with 471~ms accumulation time. The mass of the ground state, $\mathrm{ME} = -79629.6(27)$~keV, was measured against $^{133}$Cs and it is $6(4)$~keV less bound than the AME20 value \cite{AME20}. The isomer excitation energy, ($E_{x} = 139.3(20)$~keV), was measured against the ground state. Our result is in agreement with the NUBASE20 extrapolation ($E_{x,{\rm lit.}} = 150(50)$\#~keV \cite{NUBASE20}) as well as a preliminary value of 153(14)~keV reported in Ref. \cite{Kankainen2013}.

The $(8^-)$ isomer lies 115.2(20)~keV below the $(6^-)$ 254.5(3)~keV state. Thus, it can be expected that these two levels are connected by an $E2$ $\gamma$-ray transition. A dedicated $\beta$-decay study to search for this transition might be of interest from an astrophysical point of view, since the population of the isomer in $^{132}$Sn decay would change the release of energy over time \cite{Misch2021}. 

\begin{figure}
    \centering
    \includegraphics[width=\columnwidth]{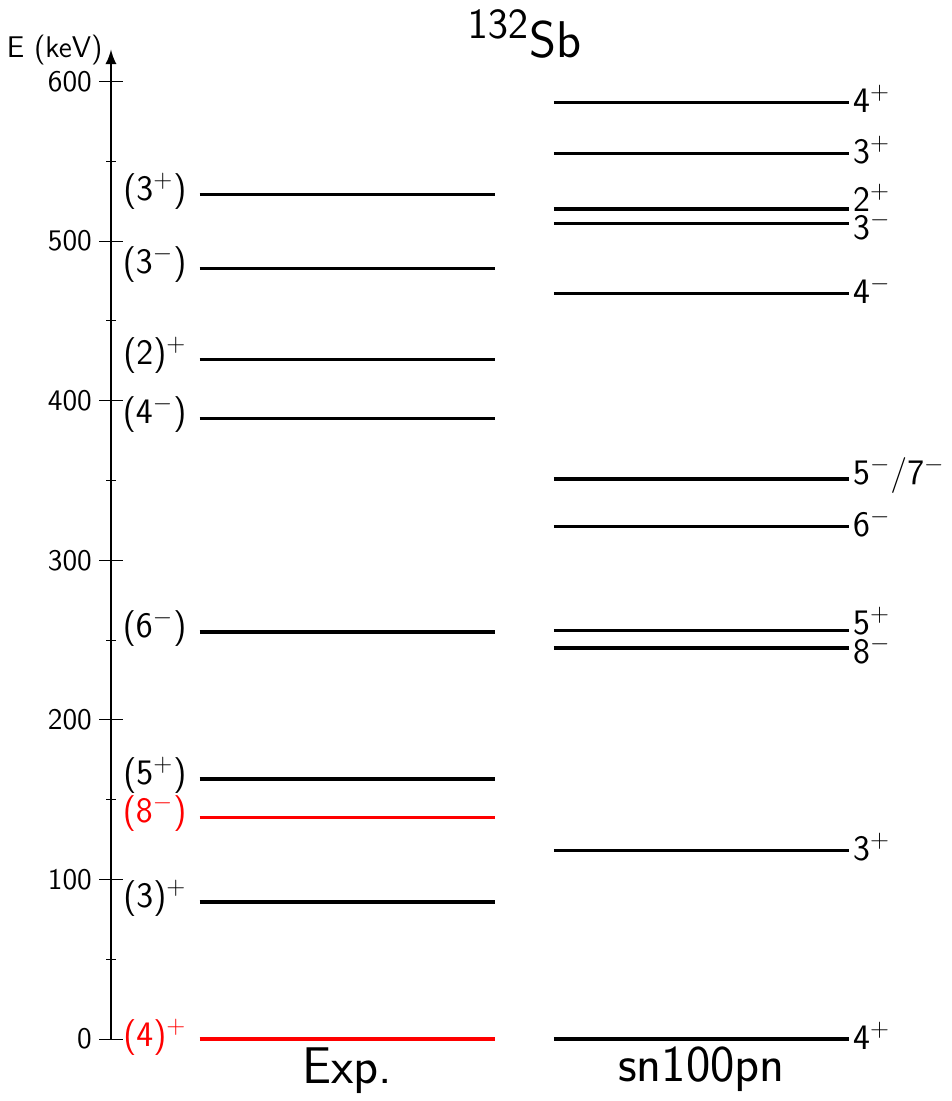}
    \caption{\label{fig:132SbSM}Comparison experimental and theoretical excited states in $^{132}$Sb up to 600 keV. The experimental data is adapted from Ref. \cite{ENSDF} based on the isomer excitation energy measured in this work, indicated in red.}
\end{figure}

Excited states in $^{132}$Sb were calculated within the shell model framework using the sn100pn interaction \cite{Brown2005} and the {\sc kshell} code. The experimental results are reproduced relatively well by the theory, see Fig. \ref{fig:132SbSM}. While the energy differences between the calculated states are slightly larger compared to the experimental data, the order is maintained and all the low-lying states are reproduced. We note that shell model predicts two states at around 350~keV, $5^-$ and $7^-$, which were not observed experimentally. However, the experimental data for this nucleus is quite limited as it was studied only via $\beta$ decay \cite{Stone1989} and spontaneous fission \cite{Bhattacharyya2001}. At the same time, the fact that the excitation energy of the $(8^-)$ isomer is reproduced relatively well benchmarks the used interaction and gives credibility to the location of these $5^-$ and $7^-$ states, not seen experimentally.

The wave function of the lowest-lying positive-parity states, including $3^+$ and $4^+$ have, as the leading component, the proton $0g_{7/2}$ orbital and the neutron $1d_{3/2}$ orbital. In the case of the lowest-lying negative-parity states, including $7^-$ and $8^-$, the shell model has as the leading component the proton $0g_{7/2}$ orbital and the neutron $0h_{11/2}$ orbital.

\section{Conclusions}

In this work we have reported mass values of eight ground states and nine isomeric states. Many of the studied nuclei are located near closed neutron shells at $N=50$ and $N=82$ and therefore provide valuable inputs to benchmark theoretical models. For the $N=49$ nucleus $^{84}$Br, we have improved the precision of the isomeric-state excitation energy by almost two orders of magnitude. The shell-model calculations with the jun45 interaction correctly obtain $2^-$ as the ground-state spin, however, the isomer is predicted to be $7^-$ and not $(6^-)$ as evaluated in NUBASE20. Further experiments, to verify the spin-parity of the isomer would be welcome in future, e.g. via laser spectroscopy. 

We have also determined five isomeric states in the region close to doubly magic $^{132}$Sn and tested the performance of theoretical models in predicting their energies and spin-parities. The excitation energy of the $(8^-)$ isomer in $^{132}$Sb, $E_x = 139.3(20)$~keV, was determined for the first time, and the excitation energies of the $1/2^-$ isomers in $^{127,129}$In were measured with a much better precision and found to agree with the NUBASE20 values \cite{NUBASE20}. The excitation energies for the high-spin ($21/2^-$) and ($23/2^-$) states were measured for the first time with a few-keV precision, and were found to be lower than in literature by $53(40)$~keV and $8(50)$~keV, respectively. 

The shell-model calculations performed with the sn100pn or SNET interactions reproduced the experimental levels reasonably well for the studied nuclei close to $^{132}$Sn. While the $1/2^-$ isomers in $^{127,129}$In are better produced with the shell-model calculations using the SNET interaction than with the MQPM calculations, the MQPM approach performs better for the high-spin isomers. The order of the states is correctly predicted and the energies are closer to the experimental values. 

For two measured cases, $^{105}$Mo and $^{121}$Pd, no long-lived isomers were observed. For $^{119}$Pd, the isomer excitation energy, $E_x = 199.1(30)$~keV, was determined for the first time. The ground state was found to be more bound than the AME20 value, which matched with the isomeric-state mass determined in this work. With the new ground-state mass of $^{119}$Pd, the kink in the two-neutron separation energies at $N=73-75$, which might have been an indication of a structural change, is removed, and the values continue to decrease linearly. Thus, the binding energies do not suggest any drastic changes in the ground-state structure for the Pd isotopes in this region.

Similarly to $^{119}$Pd, the ground state of $^{122}$Ag was found to be more bound than in AME20. Only two long-lived states were observed in $^{122}$Ag. The isomeric state was resolved from the ground state and its excitation energy, $E_x = 303.7(50)$~keV, was determined for the first time. The analysis of the collected data and the literature point to only two long-lived states and that the $(3^+)$ state listed in the NUBASE20 and ENSDF evaluations \cite{NUBASE20,ENSDF} is just an artefact.

High-precision mass measurements with the PI-ICR technique offer a way to resolve isomers from the ground states and to determine isomeric excitation energies with a few keV precision. The high resolving power enables more accurate ground-state measurements as demonstrated in this work for $^{119}$Pd and $^{122}$Ag which were both found to be more bound than the previous measurements performed with a mixture of states. More accurate ground-state masses are essential for nuclear structure and astrophysics studies. The Pd isotopes studied in this work indicate that artefacts can appear in the systematics of binding energies if isomeric states are not correctly accounted for. The possibility to determine excitation energies for isomeric states via Penning-trap mass spectrometry has yielded substantial new data on isomers that can be used to benchmark theoretical models as shown here for $^{84}$Br, $^{132}$Sb, and $^{127,129}$In in the vicinity of $N=50$ and $N=82$ shell closures. Isomeric states can also play a role in astrophysics. The impact of the newly studied isomers on the astrophysical r process can help to further these investigations in the future.
\\

\begin{acknowledgments}
This project has received funding from the European Union’s Horizon 2020 research and innovation programme under Grant Agreements No. 771036 (ERC CoG MAIDEN) and No. 861198–LISA–H2020-MSCA-ITN-2019, from the European Union’s Horizon Europe Research and Innovation Programme under Grant Agreement No. 101057511 (EURO-LABS) and from the Academy of Finland projects No. 295207, 306980, 327629, 354589 and 354968. J.R. acknowledges financial support from the Vilho, Yrj\"o and Kalle V\"ais\"al\"a Foundation. D.Ku. acknowledges the support from DAAD grant number 57610603. P.C.S. acknowledges a research grant from SERB (India), No. CRG/2022/005167.
\end{acknowledgments}

\bibliographystyle{apsrev}
\bibliography{bibfile}

\end{document}